\UseRawInputEncoding

\documentclass[11pt]{article}
\usepackage{longtable,supertabular}
\usepackage{amsmath}
\usepackage{amssymb}
\usepackage{latexsym}

\title{\bf Explicit Good Subspace-metric Codes and Subset-metric Codes}
\author{Hao Chen
  \thanks{Hao Chen is with the College of Information Science and Technology/Cyber Security, Jinan University, Guangzhou, Guangdong Province, 510632, China, haochen@jnu.edu.cn. The research  of Hao Chen was supported by NSFC Grant 62032009.}}

\begin{document}

\maketitle
\begin{abstract}

In this paper motivated from subspace coding we introduce subspace-metric codes and subset-metric codes. These are coordinate-position independent pseudometrics and suitable for the folded codes. The half-Singleton upper bounds for linear subspace-metric codes and linear subset-metric codes are proved. Subspace distances and subset distances of codes are natural lower bounds for insdel distances of codes, and then can be used to lower bound the insertion-deletion error-correcting capabilities of codes. Our subspace-metric codes or subset-metric codes  can be used to construct explicit well-structured insertion-deletion codes directly. $k$-deletion correcting codes with rate approaching $1$ can be constructed from subspace codes. By analysing the subset distances of folded codes from evaluation codes of linear mappings, we prove that they have high subset distances and then are explicit good insertion-deletion codes.\\

\end{abstract}

\section{Introduction}

For a vector ${\bf a} \in {\bf F}_q^n$, the Hamming weight $wt({\bf a})$ of ${\bf a}$ is the number of non-zero coordinate positions. The Hamming distance $d_H({\bf a}, {\bf b})$ between two vectors ${\bf a}$ and ${\bf b}$ is defined to be the Hamming weight of ${\bf a}-{\bf b}$. For a code ${\bf C} \subset {\bf F}_q^n$ of dimension $k$, its Hamming distance  $d_H$ is the minimum of Hamming distances $d_H({\bf a}, {\bf b})$ between any two different codewords ${\bf a}$ and ${\bf b}$ in ${\bf C}$.  The famous Singleton bound $|{\bf C}|\leq q^{n-d_H+1}$ is the basic upper bound for  error-correcting codes.\\

The insdel distance $d_{insdel}({\bf a}, {\bf b})$ between two vectors ${\bf a}$ and ${\bf b}$ in ${\bf F}_q^n$ is the number of insertions and deletions which are needed to transform ${\bf a}$ into ${\bf b}$. Actually it was proved in  \cite{HS17} that $$d_{insdel}({\bf a}, {\bf b})=2(n-l),$$ where $l$ is the length of the longest common subsequence or substring of ${\bf a}$ and ${\bf b}$. For two strings  ${\bf a} \in {\bf F}_q^n$ and ${\bf b} \in {\bf F}_q^m$ of different lengths the insdel distance can be defined similarly by $d_{insdel}({\bf a},{\bf b})=m+n-2l$ where $l$ is the length of the longest common substring of ${\bf a}$ and ${\bf b}$. This insdel distance $d_{insdel}$ is indeed a metric on ${\bf F}_q^n$. It is clear $$d_{insdel}({\bf a}, {\bf b})) \leq 2d_H({\bf a}, {\bf b})$$ since  $l \geq n-d_H({\bf a}, {\bf b})$ is valid for arbitrary two different vectors ${\bf a}$ and ${\bf b}$ in ${\bf F}_q^n$. The insdel distance of a code ${\bf C} \subset {\bf F}_q^n$ is the minimum of the insdel distances of two different codewords in this code. Hence the Singleton upper bound $$|{\bf C}| \leq q^{n-\frac{d_{insdel}}{2}+1}$$  follows from the Singleton bound for codes in the Hamming metric directly,  see \cite{HS17}. The relative insdel distance is defined as $\delta=\frac{d_{insdel}}{2n}$ since $d_{insdel}$ takes non-negative integers up to $2n$. From the Singleton bound $|{\bf C}| \leq q^{n-\frac{d_{insdel}}{2}+1}$ it follows immediately $$R+\delta \leq 1.$$ For insertion-deletion codes the ordering of coordinate positions strongly affects the insdel distances of codes.\\

It has been a long-standing difficult problem to deal efficiently with synchronization errors, i.e., insertion and deletion errors, see \cite{L65,L2002,VT,L1992,AGFC,DM,DA,NJAS,KLO,SZ99,BGZ,GS,SRB,SB,BGZ,Duc21,LT21,GH21}. For the recent breakthroughs and constructions we refer to \cite{HS17,HS18,HSS18,CJLW18,CGHL21,GS,SSBD,SWGY,SRB,SB,SWWY,LSWY,CJLW18,GHL21,CSI} and a nice latest survey \cite{HS21}.  We refer to \cite{SZ99,GW17,HS17,CGHL21} for asymptotic results about the rate-distance tradeoff of insertion-deletion codes. The asymptotic half-Singleton bound for linear insertion-deletion codes was proved in Section 5 in \cite{CGHL21}. It can be restated as $$d_{insdel} \leq \max\{2(n-2k+2),2\}$$ for the insdel distance of  a  linear $[n,k]_q$ insertion and deletion code, see  \cite{Chen21}. This half-Singleton bound for linear insertion-deletion codes can be generalized to the strong half-Singleton bound based on the generalized Hamming weights $$d_{insdel} \leq 2(d_r-2r+2),$$ where $d_1,\ldots, d_k$ are generalized Hamming weights, we refer to \cite{Chen21}.\\

The problem to construct explicit $k$-deletion correcting codes of rate approaching $1$ as length grows had been unsettled for a long time. The Varshamov-Tenengolts code $${\bf VT}=\{{\bf c}: \Sigma_{i=1}^n ic_i \equiv 0 mod (n+1)\}$$ was proposed in \cite{VT} and generalized in \cite{HF} by the Fibonacci weights and further generalized to number-theoretic codes in \cite{PAGFC}.  There are very few known systemic construction of explicit insertion-deletion codes for various lengths, cardinalities and insdel distances. Most insertion-deletion codes in \cite{GW17, HS17,HS18,HS21,CSI} have been only given algorithmically. They are not explicit codes, though sometimes these nice insertion-deletion codes can be constructed from highly efficient polynomial time algorithms. On the other hand $k$-deletion correcting codes with optimal redundancies developed in \cite{VT,T84,HF,KLO,BGZ,GS,SRB,SB,GH21} have been explicitly given.  From both theoretical and practical motivations systemic constructions of explicit good well-structured insertion-deletion codes are necessary and important. A direct method to obtain explicit insertion-deletion code is the position-indexing Hamming error-correcting codes. As showed in this paper from subspace-metric and subset metric codes introduced in this paper, many explicit near-optimal subspace-metric and subset-metric, then insertion-deletion codes can be constructed directly and systemically.\\

Subspace subcodes are the subcodes of codes over ${\bf F}_{q^n}$ consisting of codewords with coordinates in a fixed ${\bf F}_q$ linear subspace of ${\bf F}_{q^n}$. It was first considered in \cite{HMS} for Reed-Solomon codes, as codes over smaller alphabets. It was showed that these subspace subcodes of Reed-Solomon codes have quite nice coding parameters. Then the generalized subspace subcodes was considered in \cite{BGK}.  In this paper we introduce the subspace-metric and the subset-metric on  ${\bf F}_{q^n}^m$ which measures the distances of subspaces spanned by coordinates of codewords, and the distances of the subsets consisting of coordinates of codewords. These are pseudometrics satisfying the following property.\\
1) $d(x,y) \geq 0$;\\
2) $d(x,y)=d(y,x)$;\\
3) $d(x,z) \leq d(x,y)+d(y,z)$.\\
However it is possible that $d(x,y)=0$ for some different elements $x$ and $y$.  These two pseudometrics are coordinate-position independent. \\

Folded codes was introduced in \cite{GR08} to achieve the list-decoding capacity. These codes have nice list-decoding properties. For a code ${\bf C}$ in ${\bf F}_q^{rs}$, the folded code with the parameter $s$ is the code $Folded({\bf C}) \subset {\bf F}_{q^s}^r$ consisting of $({\bf c}_1, \ldots,{\bf c}_r)$, where ${\bf c}=(c_1,\ldots,c_{rs}) \in {\bf C}$, and ${\bf c}_i=(c_{(i-1)s+1},\ldots,c_{is})$ for $i=1,2,\ldots,r$.  The subspace-metric and subset-metric are suitable measures for folded codes. We will prove that folded codes from some evaluation codes of linear mappings have relative high subset distances. Then they have relative high insertion-deletion error-correcting capabilities.\\бо

Subspace-metric and subset-metric are on the one hand lower bounds for the insdel metric, on the other hand have their own interests. Comparing with the rank-metric subspace-metric can be considered as coordinate position independent version of rank-metric. For example if there exist  two codewords of the shapes $(c_1, \ldots, c_n)$ and $(c_2,\ldots,c_n,c_1)$ of a code ${\bf C} \subset {\bf F}_{q^m}^n$, the subspace distance of this code is zero, on the other hand the rank distance might be large.  Hence it is not unnatural to introduce subspace-metric and subset-metric, which are coordinate position independent pseudometrics, as metrics between rank-metric and insdel metric. It seems hopeful that these two pseudometrics have other applications besides lower bounding insdel distances.\\

\section{Subspace-metric  and subset-metric codes}

For two vectors ${\bf x}=(x_1,\ldots,x_m) \in {\bf F}_{q^n}^m$ and ${\bf y}=(y_1,\ldots,y_m) \in {\bf F}_{q^n}^m$, let $S_{{\bf x}}=<x_1,\ldots,x_m>$  and $S_{{\bf y}}=<y_1,\ldots,y_m>$ be the two ${\bf F}_q$ linear subspaces in ${\bf F}_{q^n}={\bf F}_q^n$ spanned by $x_1,\ldots,x_m$, and  $y_1,\ldots,y_m$. The subspace distance between these two vectors is $$ d_S({\bf x},{\bf y})=\dim (S_{{\bf x}}+S_{{\bf y}})-\dim (S_{{\bf x}} \bigcap S_{{\bf y}})$$ The subset distance between these two vectors is $$d_{subset}({\bf x},{\bf y}) =|\{x_1,\ldots,x_m\} |+|\{y_1,\ldots,y_m\}|-2|\{x_1,\ldots,x_m\} \bigcap \{y_1,\ldots,y_m\}|.$$ It can be verified  $$d_S({\bf x}, {\bf y}) \leq d_{subset}({\bf x}, {\bf y}) \leq d_{insdel}({\bf x},{\bf y}) \leq 2d_H({\bf x}, {\bf y}), $$ we refer to Section 3.\\

{\bf Lemma 2.1.} {\em We have $d_S({\bf x}, {\bf z}) \leq d_S({\bf x},{\bf y})+d_S({\bf y},{\bf z})$ and $d_{subset}({\bf x}, {\bf z}) \leq d_{subset}({\bf x},{\bf y})+d_{subset}({\bf y},{\bf z})$.}\\

Then the subspace distance and subset distance are indeed pseudometrics on ${\bf F}_{q^n}^m$. For a code ${\bf C} \subset {\bf F}_{q^n}^m$ the minimum subspace distance and the minimum subset distance are the minimum subspace distance and minimum subset distances between its codewords,  $$d_S({\bf C})=\min_{{\bf x} \neq {\bf y}}\{d_S({\bf x}, {\bf y}):{\bf x}, {\bf y} \in {\bf C}\},$$ $$d_{subset}({\bf C})=\min_{{\bf x} \neq {\bf y}}\{d_{subset}({\bf x}, {\bf y}):{\bf x}, {\bf y} \in {\bf C}\}.$$
 Then $d_S({\bf C})) \leq d_{subset}({\bf C}) \leq 2d_H({\bf C})$ from Theorem 3.1. When a subspace-metric or subset metric code ${\bf C}$ is linear, then  $d_S({\bf C})\leq d_H({\bf C})$ and $d_{subset}({\bf C}) \leq d_H({\bf C})$.\\

Because of their property of being pseudometric, in the case that the linear span $<x_1,\ldots,x_m>={\bf F}_{q^n}$ is the whole space ${\bf F}_q^n$ for many codewords ${\bf x} \in {\bf C}$, the minimum subspace distance of this code would be trivial. Hence we introduce the following $r$-th subspace distance between two vectors in ${\bf F}_{q^n}^m$ where $m$ is a positive integer which is divisible by $r$. It is clear these $r$-th subspace distance and the $r$-th subset distance are just the subspace distance and the subset distance of the folded code introduced  in \cite{GR08}.\\

Let $m=rg$ where $r$ and $g$ are positive integers. For a  vector ${\bf x}=(x_1,\ldots,x_{rg}) \in {\bf F}_{q^n}^{rg}$  set ${\bf x}_1=(x_1,\ldots,x_r)$, $\ldots, {\bf x}_g=(x_{(g-1)r+1},\ldots,x_{gr})$. For two vectors  ${\bf x}$ and ${\bf y}$ in ${\bf F}_{q^n}^{rg}$. Set $$S_{r, {\bf x}}=<{\bf x}_1, \ldots, {\bf x}_g>$$  and $$S_{r, {\bf y}}=<{\bf y}_1, \ldots,{\bf y}_g>$$ be the two ${\bf F}_q$ linear subspaces in ${\bf F}_{q^n}^r={\bf F}_q^{nr}$ spanned by $g$ vectors ${\bf x}_1,\ldots,{\bf x}_g$ and $g$ vectors ${\bf y}_1,\ldots, {\bf y}_g $ in ${\bf F}_q^{nr}$.  The $r$-th subspace distance between these two vectors ${\bf x}$ and ${\bf y}$ is $$ d_{r,S}({\bf x},{\bf y})=\dim (S_{r, {\bf x}}+S_{r, {\bf y}})-\dim (S_{r, {\bf x}} \bigcap S_{r, {\bf y}})$$ The $r$-th subset distance between ${\bf x}$ and ${\bf y}$ is $$d_{r,subset}({\bf x},{\bf y})=|\{{\bf x}_1, \ldots, {\bf x}_g\}|+|\{{\bf y}_1, \ldots, {\bf y}_g\}|-2|\{{\bf x}_1, \ldots, {\bf x}_g\} \bigcap \{{\bf y}_1, \ldots, {\bf y}_g\}|.$$  It can be verified that $d_{r,S}$ and $d_{r,subset}$  are also pseudometrics on ${\bf F}_{q^n}^{gr}$.\\

The minimum $r$-th subspace distance $d_{r,S}({\bf C})$ and the $r$-th subset distance of a code ${\bf C} \subset {\bf F}_{q^n}^{gr}$ are defined to be the minimum of all $r$-th subspace distances and all $r$-th subset distances of two different codewords in the code ${\bf C}$,  $$d_{r,S}({\bf C})=\min_{{\bf x} \neq {\bf y}}\{d_{r,S} ({\bf x}, {\bf y}):{\bf x}, {\bf y} \in {\bf C}\},$$  $$d_{r,subset}({\bf C})=\min_{{\bf x} \neq {\bf y}}\{d_{r,subset} ({\bf x}, {\bf y}):{\bf x}, {\bf y} \in {\bf C}\}.$$  If $r$ is not a factor of $m$ we can define the $r$-th subspace distance and $r$-th subset distance by padding some coordinates. The detail is omitted.\\

From the obvious fact $d_{r,S} ({\bf C}) \leq d_{subset}({\bf C})  \leq 2d_H({\bf C})$ we have the following Singleton bound for subspace-metric codes over ${\bf F}_q$, $$|{\bf C}|\leq q^{n-\frac{d_{r,S}}{2}+1}, $$ $$|{\bf C}|\leq q^{n-\frac{d_{subset}}{2}+1}. $$ The  half-Singleton bounds in Theorem 3.2 for linear subspace-metric codes and linear  subset-metric codes are similar to the half-Singleton bound for linear insertion-deletion codes, see \cite{CGHL21,Chen21}.\\

We consider the following trivial examples of linear subspace-metric codes. Let ${\bf C}={\bf F}_{q^n}^m$, this is a $m$ dimension linear $[m,m]_{q^n}$ code with the trivial subspace distance $d=0$. Then we consider the  linear $[mr,m]_{q^n}$ repetition-code ${\bf C}=\{({\bf x},\ldots,{\bf x}):{\bf x} \in {\bf F}_{q^n}^m\}$. Its 1st subspace distance is $0$. Its $r$-th subspace distance is $2$.  Concatenation codes from subspace-metric codes have a lower bound on their $r$-th subspace distances from the subspace distances of their outer codes.\\

It is obvious that the subspace-metric is different to the rank-metric in \cite{Gabidulin,YH4} and the sum-metric in \cite{MK}. These metrics are coordinate-position dependent and can not be used to lower bound the insdel distances of codes. In general it is difficult to give linear subspace-metric or subset -metric codes with high subspace distances. In this paper we give a highly non-trivial lower bound for the subset distances of the folded codes from evaluation codes of linear mappings.\\

\section{Lower bounds and the half-Singleton bounds}

How to  lower bound the insdel distances of codes is a very challenge problem, since the accurate positions of the common substring of two codewords are different. In this paper motivated from subspace coding we introduce subspace-metric and subset-metric codes.  These two pseudometrics are coordinate-position independent. Subspace-metric is different to the rank-metric and has its own interest. Then the half-Singleton bounds for the linear subspace-metric and linear subset-metric codes are proved. The subspace distance and the subset distance are natural lower bounds for the insdel distances. Then codes with high subspace distances or high subset distances have high insertion-deletion error-correcting capababilities.  From explicitly constructed subspace codes we can get many explicit good subspace-metric and subset-metric codes, then explicit good insertion-deletion codes. \\

The following Theorem 3.1  gives a way to lower bound the insdel distances from the subspace distances and the subset distances. \\

{\bf Theorem 3.1  (lower bounds).} {\em Let ${\bf C}$ be a code in ${\bf F}_{q^n}^m$. Then we have $$d_{insdel}({\bf C}) \geq \min_{{\bf x}\neq {\bf y}}\{d_{subset}({\bf x}, {\bf y})\} \geq \min_{{\bf x}\neq {\bf y}}\{d_S({\bf x}, {\bf y})\}.$$ }\\

{\bf Proof.}  We prove the lower bound on the insdel distance from the 1st subspace distance. For two different codewords ${\bf x}$ and ${\bf y}$ in ${\bf C}$, there are at least $$\dim (S_{{\bf x}})-\dim(S_{{\bf x}}\bigcap S_{{\bf y}})$$ coordinates in the set $\{x_1,\ldots,x_m\}$ not in the intersection $\{x_1,\ldots,x_m\} \bigcap \{y_1,\ldots,y_m\}$. Similarly there are at least $$\dim (S_{{\bf y}})-\dim(S_{{\bf x}}\bigcap S_{{\bf y}})$$ coordinates in the set $\{y_1,\ldots,y_m\}$ not in the intersection $\{x_1,\ldots,x_m\} \bigcap \{y_1,\ldots,y_m\}$. Then there are at least $$\dim (S_{{\bf x}})+\dim (S_{{\bf x}})-2\dim(S_{{\bf x}}\bigcap S_{{\bf y}})$$ different coordinates in the set $\{x_1,\ldots,x_m\}$ and $\{y_1,\ldots,y_m\}$.  The case of $r$-th subspace distance lower bound can be proved similarly. \\

For two different codewords ${\bf x}=(x_1, \ldots, x_n ) \in {\bf F}_{q^m}^n$ and ${\bf y}=(y_1, \ldots, y_n) \in  {\bf F}_{q^m}^n$. Let $\{x_{i_1}, \ldots, x_{i_t}\}$ be the set $\{x_1, \ldots,x_n\} -\{x_1, \ldots, x_n\} \bigcap \{y_1, \ldots, y_n\}$ and $\{y_{j_1}, \ldots, y_{j_h}\}$ be the set $\{y_1, \ldots,y_n\} -\{x_1, \ldots, x_n\} \bigcap \{y_1, \ldots, y_n\}$. Set $S_1$ the linear subspace in ${\bf F}_{q^m}={\bf F}_q^m$ spanned by coordinates in $\{x_{i_1}, \ldots, x_{i_t}\}$ and $S_2$ the linear subspace  in ${\bf F}_{q^m}={\bf F}_q^m$ spanned by coordinates in $\{y_{j_1}, \ldots, y_{j_t}\}$. Then $d_S({\bf x}, {\bf y}) \leq \dim(S_1)+\dim(S_2)-\dim (S_1 \bigcap S_2) \leq d_{subset}({\bf x}, {\bf y})=|\{x_{i_1}, \ldots, x_{i_t}\}|+|\{y_{j_1}, \ldots, y_{j_h}\}|$. The conclusion is proved.\\

We give the half-Singleton bounds for linear subspace-metric and linear subset-metric codes.\\

{\bf Theorem 3.2.} {\em Let ${\bf C} \subset {\bf F}_{q^m}^n$ be a linear $[n, k]_q$ code with the minimum subspace distance $d_S({\bf C})$ and the minimum subset distance $d_{subset}({\bf C})$. If the rate of this code is bigger than $\frac{1}{2}$. that is, $k >\frac{n}{2}$, then $$d_S({\bf C})=d_{subset}({\bf C})=0.$$  Moreover we have $$d_S({\bf C}) \leq d_{subset}({\bf C}) \leq \max\{2(n-2k+2),2\}.$$ Based on the generalized Hamming weights we have $$d_S({\bf C}) \leq d_{subset}({\bf C}) \leq \min_{1 \leq r \leq k} \{d_r({\bf C})-2r+2\},$$ where $d_1({\bf C}),d_2({\bf C}),\ldots,d_k({\bf C})$ are the generalized Hamming weights of this code ${\bf C}$.}\\

{\bf Proof.} We prove the following claim. If $k >\frac{n}{2}$, then there exists a non-zero codeword ${\bf x}=(x_1,\ldots,x_n) \in {\bf C}$ such that ${\bf y}=(x_2,x_3,\ldots,x_n,x_1)$ is also a codeword in ${\bf C}$. Let ${\bf H}$ be the $(n-k) \times n$ parity-check matrix of this code ${\bf C}$ with $n$ columns ${\bf h}_1,\ldots,{\bf h}_n$. We form two new matrices as follows. One is the $(n-k) \times n$ matrix ${\bf H}'=({\bf h}_2,{\bf h}_3,\ldots,{\bf h}_n,{\bf h}_1)$. Another is the $2(n-k) \times n$ matrix ${\bf H}''$ by concatenation corresponding columns in ${\bf H}$ and ${\bf H}'$, that is, the $n$ columns in ${\bf H}''$ are $n$ vectors $$({\bf h}_1,{\bf h}_2)^{\tau},({\bf h}_2,{\bf h}_3)^{\tau},\ldots,({\bf h}_{n-1},{\bf h}_n)^{\tau}, ({\bf h}_n, {\bf h}_1)^{\tau}$$ in ${\bf F}_q^{2(n-k)}$. Since $2(n-k)<n$, there is a non-zero solution of the equation $${\bf H}'' \cdot {\bf x}^{\tau}={\bf 0}.$$ Then $d_S({\bf x}, {\bf y})=d_{subset}({\bf x}, {\bf y})=0$. . Hence $d_S({\bf C})=d_{subset}({\bf C})=0$. The second conclusion follows from Theorem 3.1 and the half-Singleton bound for the insdel distance in \cite{CGHL21,Chen21}. The conclusion is proved.\\

\section{Subspace codes}

Subspace codes have been studied extensively since the paper \cite{KK} of R. K\"{o}tter and F. R. Kschischang. It was proposed to correct errors and erasures in network transmissions of information. A set ${\bf C}$ of $M$ subspaces of the dimension $k \in T$ in ${\bf F}_q^n$, where $T$ is a subset of $\{1,2,\ldots, n-1\}$,  is called an $(n, M, d, T)_q$ subspace code if  $d_S(U,V)=\dim U+\dim V-2\dim(U \cap V) \geq d$ is satisfied for any two different subspaces $U,V$ in ${\bf C}$. The main problem of the subspace coding is to determine the maximal possible size ${\bf A}_q(n, d, T)$ of such a code for given parameters $n,d,T,q$. When $T$ is the whole set $\{1,2,\ldots,n\}$, we write ${\bf A}_q(n,d)$ for the maximal possible size of the set of subspaces in ${\bf F}_q^n$ such that the subspace distances between any different subspaces in this set are at least $d$. Let $\displaystyle{n \choose k}_q=\prod_{i=0}^{k-1} \frac{q^{n-i}-1}{q^{k-i}-1}$ be the $q$-ary Gauss coefficient, which is the number of $k$-dimensional subspaces in ${\bf F}_q^n$. It is clear $${\bf A}_q(n, d, T) \leq \Sigma_{k \in T}  \displaystyle{n \choose k}_q$$ and $${\bf A}_q(n, d) \leq \Sigma_{k=1}^{n-1}  \displaystyle{n \choose k}_q.$$\\

When $T=\{k\}$ contains only one dimension this is a constant dimension subspace code, otherwise it is called a mixed dimension subspace code. There have been some upper and lower bounds for ${\bf A}_q(n,d,k)$. We refer to papers \cite{EtzionVardy,Silberstein1,Silberstein2,XuChen,CHWX} and the nice webpage \cite{table} for latest constructions and papers. We refer to the nice surveys \cite{Honold} and the webpage \cite{table}. \\

\subsection{Rank-metric codes}

Rank-metric codes have been widely used in the constructions of large constant dimension subspace codes. The rank-metric on the space ${\bf M}_{m \times n}({\bf F}_q)$ of size $m \times n$ matrices over ${\bf F}_q$ is defined by the rank of matrices, i.e., $d_r(A,B)= rank(A-B)$. The minimum rank-distance of a code ${\bf M} \subset {\bf M}_{m \times n}({\bf F}_q)$ is defined as $$d_r({\bf M})=\min_{A\neq B} \{d_r(A,B): A \in {\bf M}, B\in {\bf M} \}$$ For a code ${\bf M}$ in ${\bf M}_{m \times n}({\bf F}_q)$ with the minimum rank distance $d_r({\bf M}) \geq d$, it is well-known that the number of codewords in ${\bf M}$ is upper bounded by $q^{\max\{m,n\}(\min\{m,n\}-d+1)}$ , see \cite{Gabidulin}. A code attaining this bound is called a maximum rank-distance (MRD) code. \\

The Gabidulin code ${\bf Q}_{q,n,t}$ consisting of ${\bf F}_q$ linear mappings on ${\bf F}_q^n \cong {\bf F}_{q^n}$ defined by $q$-polynomials $a_0x+a_1x^q+\cdots+a_ix^{q^i}+\cdots+a_tx^{q^t}$, where $a_t,\ldots,a_0 \in {\bf F}_{q^n}$ are arbitrary elements in ${\bf F}_{q^n}$, is an MRD code, see \cite{Gabidulin}. The rank-distance of ${\bf Q}_{q,n,t}$ is $n-t$ since there are at most $q^t$ roots in ${\bf F}_{q^n}$ for each such $q$-polynomial. There are  $q^{n(t+1)}$ such $q$-polynomials in ${\bf Q}_{q,n, t}$. Let $h$ be a non-negative integer and $\phi: {\bf F}_{q^k} \longrightarrow {\bf F}_{q^{k+h}}$ be a $q$-linear embedding. Then $$a_t \phi(x^{q^t})+a_{t-1}\phi(x^{q^{t-1}})+\cdots+a_1\phi(x^q)+a_0\phi(x)$$ is a $q$-linear mapping from ${\bf F}_{q^k}$ to ${\bf F}_{q^{k+h}}$, where $a_i \in {\bf F}_{q^{k+h}}$ for $i=0,1,\ldots,t$. We denote the set of all such mappings as ${\bf Q}_{q, k\times (k+h), t}$. It is clear that the dimension of the kernel of any such mapping is at most $t$. Then ${\bf Q}_{q, k\times (k+h), t} \subset {\bf M}_{k \times (k+h)}({\bf F}_q)$ is an MRD code with rank distance $k-t$ and $q^{(k+h)(t+1)}$ elements. These MRD codes have been used widely in previous constructions of constant dimension subspace codes, see \cite{Silberstein1,Silberstein2,Honold}.\\

\subsection{Lifted rank-metric code}

Let $n$ and $m$ be two positive integers satisfying $m \geq n$. For any given rank-metric code ${\bf M}$ with the cardinality $M$ in ${\bf M}_{n \times m}({\bf F}_q)$ with the rank distance $d$, we have an $(n+m, M, 2d,n)_q$ constant dimension subspace code consisting of $M$ subspaces of dimension $n$  in ${\bf F}_q^{n+m}$ spanned  by the rows of $(I_n, A)$, where $A$ is an element in ${\bf M}$. Here $I_n$ is the $n \times n$ identity matrix. Let $U_A$ be the subspace spanned by rows in $(I_n,A)$. The intersection $U_A \cap U_B$ is the set $\{ (\alpha,\alpha A)=(\beta, \beta B): \alpha (A-B)=0, \alpha \in {\bf F}_q^n\}$. Thus $\dim(U_A \cap U_B)  \leq n-d$. The subspace distance of this constant dimension subspace code is at least $2d$. A constant dimension subspace code  constructed as above is called a lifted  rank-metric code. When ${\bf M}$ is a MRD (maximal rank distance) code we have the following result. Suppose that $n \leq m $  then $${\bf A}_q(n+m, 2d, n) \geq q^{m(n-d+1)}.$$ The corresponding $q^{m(n-d+1)}$ subspaces of dimensions $n$ in ${\bf F}_q^{n+m}$ are spanned by rows in $(I_n, A)$ where $A$ is from all elements in the Gabudilin code of linearized $q$-polynomials of the form $\Sigma_{i=0}^{n-d} a_i\phi(x^{q^i})$ where $a_i \in {\bf F}_{q^m}$.\\

\section{Explicit subspace-metric codes from subspace codes}

Explicit good subspace-metric  codes can be constructed from subspace codes naturally. From the lower bound $d_{insdel} \geq d_S$, then explicit good insertion-deletion codes can be constructed from these subspace-metric codes.\\

{\bf Theorem 5.1.} {\em Let ${\bf C}$ be a constant or mixed dimension subspace code of subspaces in ${\bf F}_q^n$ with the cardinality $M$ and the subspace distance $d$, then for any length $l\geq \max_{L \in {\bf C}}\dim(L)$ , we have a length $l$ subspace-metric  code $Span({\bf C})$ over ${\bf F}_{q^n}$ with the subspace distance $d_S$ satisfying $ d_S \geq d$ and the cardinality $M$}.\\

{\bf Proof.} The construction of span code $Span({\bf C})$ from the subspace code ${\bf C}$ is as follows. For any  subspace $L \in {\bf C}$, we take $l$ vectors ${\bf c}_1(L), \ldots, {\bf c}_l(L) $ from $L$, ${\bf c}_i (L) \in L \subset {\bf F}_q^n={\bf F}_{q^n}$, such that ${\bf c}_1(L), \ldots, {\bf c}_l(L)$ span ${\bf L}$ as a linear subspace of ${\bf F}_q^n$. Then we have one codeword ${\bf c}(L)=({\bf c}_1(L), \ldots, {\bf c}_l(L))$ in $Span({\bf C})$ from each codeword $L$ in ${\bf C}$.\\

For two different codewords ${\bf L}_1$ and ${\bf L}_2$ in the subspace codes ${\bf C}$, suppose the longest common subsequence of ${\bf c}(L_1)$ and ${\bf c}(L_2)$ has length $\mu$, then we have $$\dim(L_1) -\dim(L_1 \bigcap L_2) \leq l-\mu,$$ $$\dim(L_2) -\dim(L_1 \bigcap L_2) \leq l-\mu,$$ since the coordinates in ${\bf c}(L_1)$ and ${\bf c}(L_2)$ span the whole subspace $L_1$ and $L_2$. Hence we have $$d_{subet}(Span({\bf C}))=2(l-\mu) \geq \dim(L_1) +\dim(L_2)-2\dim(L_1 \bigcap L_2) \geq d_S({\bf C}).$$

{\bf Theorem 5.2.} {\em Let ${\bf C}$ be a constant dimension subspace code of $k$-dimensional subspaces in ${\bf F}_q^n$ with the cardinality $M$ and the subspace distance $2k-2t$, then for any length $t+1 \leq l\leq k$ , we have a length $l$ subspace-metric code $Span({\bf C})$ over ${\bf F}_{q^n}$ with the subspace distance $d_S \geq 2(l-t)$ and the cardinality $M$}.\\

{\bf Proof.} For each $k$ dimensional subspace $L$ in ${\bf C}$, we take $l$ linearly independent vectors $({\bf c}_1(L),\ldots,{\bf c}_l(L))\in {\bf F}_{q^n}^l$ in $L$. Then any two such different codewords in the insertion-deletion code $Span({\bf c})$ have at most $t$ common vectors since they are in two different subspaces in ${\bf C}$. The conclusion follows directly.\\

{\bf Corollary 5.1.} {\em If there is an explicit $(n, M, d, k)_q$ constant dimension subspace code, then for any positive integer $l$ satisfying $q^{k-\frac{d}{2}} <l \leq q^k$, we have an explicit insertion-deletion code over ${\bf F}_{q^n}$ with the length $l$, insdel distance $2(l-q^{k-\frac{d}{2}})$ and the the cardinality $M$.}\\

{\bf Proof.} We take all vectors in each $k$-dimensional subspace, then the explicit insertion-deletion code is constructed.\\

{\bf Theorem 5.3.} {\em Let ${\bf C}$ be a constant dimension subspace code of $k$-dimensional subspaces in ${\bf F}_q^n$ with the cardinality $M$ and the subspace distance $d$, then we have a length $kn$ insertion-deletion code $Insdel({\bf C})$ over ${\bf F}_q$ with insdel distance $d$ and the cardinality $M$}.\\

{\bf Proof.}  We take $k$ linearly independent vectors $({\bf c}_1(L),\ldots,{\bf c}_k(L))\in {\bf F}_{q^n}^k$ in $L$ for each $k$ dimensional subspace $L$ in ${\bf C}$. Fixed a basis of ${\bf F}_{q^n}$ over ${\bf F}_q$ and $exp({\bf x})$ is the expansion with respect to this basis for ${\bf x} \in {\bf F}_{q^n}$.  Then this vector $({\bf c}_1(L),\ldots,{\bf c}_k(L))$ in ${\bf F}_{q^n}^k$ is considered as a vector $(exp({\bf c}_1(L)), \ldots, exp({\bf c}_k(L))) \in {\bf F}_q^{kn}$ as a codeword in the insertion-deletion code $Insdel({\bf C}) \subset {\bf F}_q^{nk}$. Suppose any given two such different codewords  $(exp({\bf c}_1(L)), \ldots, exp({\bf c}_k(L)))$ and  $(exp({\bf c}_1(L')), \ldots, exp({\bf c}_k(L')))$ in $Insdel({\bf C})$ have a length $T$ common subsequence, then $nk-T \geq \frac{d}{2}$. Otherwise be deleting $nk-T <\frac{d}{2}$ basis vectors of $L$ and $L'$, the remaining basis vectors of $L$ and $L'$ have to be the same. This is a contradiction to the fact $d_S(L.L') \geq d$. The conclusion follows directly.\\

From Theorem 5.1, Theorem 5.2, Corollary 5.1 and Theorem 5.3 many good explicit insertion-deletion codes can be constructed via the known good subspace codes in \cite{table}.\\

In this section we give several explicit subspace-metric codes with their cardinalities close to the Singleton bound.\\

{\bf Theorem 5.4.} {\em From the lifted MRD constant dimension subspace $(2n, q^{n(t+1)}, \\2(n-t), n)_q$ code  ${\bf C}$ we have a length $n$ subspace-metric  code $Span({\bf C})$ over ${\bf F}_{q^{2n}}$ with the relative subspace distance $\delta$ and the rate $\frac{1-\delta}{2}$.}\\

{\bf Proof.} From Theorem 5.1 we have a length $n$ subspace-metric code over ${\bf F}_{q^{2n}}$ with the subspace distance $2(n-t)$ and the cardinality $q^{n(t+1)}$ directly. Then the conclusion follows.\\

Actually from the results in \cite{XuChen,CHWX} new $n$-dimensional subspaces can be added to the above lifted MRD subspace codes preserving the subspace distances $2(n-t)$.  For example we have the following result. From Theorem 5.1 further results about larger subspace-metric codes can be obtained by the constant dimension subspace codes in \cite{XuChen,CHWX}.\\

{\bf Corollary 5.2.} {\em Let $t $ be a positive integer satisfying $t \geq \frac{n}{2}$. We have a length $n$ subspace-metric code over ${\bf F}_{q^{2n}}$ with the subspace  distance $2(n-t)$ and the cardinality $$q^{n(t+1)}+\Sigma_{i=n-t}^{t} {\bf rank}_i({\bf Q}_{q,n,t}).$$ Without the condition $t \geq \frac{n}{2}$ we have a length $n$ subspace-metric code over ${\bf F}_{q^{2n}}$ with the subspace distance $2(n-t)$ and the cardinality $q^{n(t+1)}$.}\\

{\bf Corollary 5.3.} {\em Let $t$ be a positive integer satisfying $t\geq \frac{n}{2}$ and $s$ be an arbitrary positive integer. Then a length $n$ subspace-metric code over the large field ${\bf F}_{q^{(s+1)n}}$ with the subspace distance $2(n-t)$ and the cardinality $\Sigma_{j=0}^s q^{(s-j)n(t+1)}(\Sigma_{i=t}^{n-t}{\bf rank}_i({\bf Q}_{q,n,t}))^j$ can be constructed from the insdel-subspace coding connection. Without the condition $t \geq \frac{n}{2}$ we have a length $n$ subspace-metric code over ${\bf F}_{q^{(s+1)n}}$ with the subspace distance $2(n-t)$ and the cardinality $q^{sn(t+1)}$.}\\

Then from Theorem 3.1  we have an insertion-deletion code over ${\bf F}_{q^{(s+1)n}}$ with the relative insdel distance $1-\delta$ and the rate $\frac{s}{s+1}\delta$ for any given positive real number $\delta <1$. When $s$ goes to the infinity this is a near-Singleton bound insertion-deletion code over very large fields.  The Singleton bound of an insertion-deletion code in ${\bf F}_{q^{(s+1)n}}^n$ with the insdel distance $2(n-t)$ is $q^{(s+1)n(t+1)}$.\\

Actually the subspace-metric code in Theorem 5.4 can be better. The key point is as follows. We will use not only $n$ linearly independent vectors in each $n$-dimensional subspace in a constant dimension subspace code, more codewords in each subspace in this constant dimension subspace code will be added to this subspace-metric code.\\

The $n$ linear independent vectors in each subspace of an $(2n, q^{n(t+1)}, 2(n-t), n)_q$ lifted MRD constant dimension subspace code can be taken from $n$ rows in the $n \times 2n$ matrix $(I_n,A)$ where $A$ takes all codewords from the Gabidulin rank-metric code of rank distance $n-t$. Here we take more $n$ linear independent rows of the $n \times 2n$ matrix $(G,GA)$ where $G$ is an $n \times n$ non-singular matrix and $A$ takes all codewords from the Gabidulin rank-metric code of rank-distance $n-t$.  If $A$ and $B$ are different codewords in the Gabidulin rank-metric code of the rank distance $n-t$, it is clear the subspace distance between $n$ rows of $(G_1,G_1A)$ and $(G_2,G_2B)$ is at least $2(n-t)$ because they span different $n$-dimensional subspaces in this $(2n, q^{n(t+1)}, 2(n-t), n)_q$ lifted MRD constant dimension subspace code. Then the key point is that how many non-singular $n \times n$ matrices $G$ can be taken such that the subspace distance between $n$ rows of $(G_1,G_1A)$ and $(G_2,G_2A)$ is at least $2(n-t)$. Here we require that there are at most $t$ common rows for any two given non-singular $n \times n$ matrices $G_1$ and $G_2$. Then it follows that there are at most $t$ common rows of the two $n \times 2n$ matrices $(G_1,G_1A)$ and $(G_2,G_2A)$.\\

One construction is as follows. We set $G$ as the following form.\\

$$
\left(
\begin{array}{ccccc}
I_{\frac{n}{2}}&H_1\\
0&H_2\\
\end{array}
\right)
$$
where $H_1$ is a $\frac{n}{2} \times \frac{n}{2}$ matrix and $H_2$ is a non-singular  $\frac{n}{2} \times \frac{n}{2}$ matrix.\\

Let $t$ be a positive integer satisfying $t \geq \frac{n}{2}$. We take $H_2$ as the multiplication of elements in ${\bf F}_{q^{\frac{n}{2}}}$. Hence we have to divide $(\frac{n}{2})^2$ elements such that the corresponding $\frac{n}{2} \times \frac{n}{2}$ matrices have no common row. Actually for fixed basis of ${\bf F}_{q^{\frac{n}{2}}}$ $e_1,\ldots,e_{\frac{n}{2}}$, we have $xe_i=ye_j$ for some different indices in $\{1,\ldots,\frac{n}{2}\}$ if the corresponding matrices of $x$ and $y$ have a common row. Hence there are $\frac{q^{\frac{n}{2}}-1}{\frac{n}{2}^2}$ possibilities for such $H_2$. On the other hand we can take $H_1$ in a Gabidulin code ${\bf Q}_{q,\frac{n}{2},t-\frac{n}{2}}$. There are $q^{\frac{n}{2}(t+1-\frac{n}{2})}$ possibilities for such $H_1$.  This process can be continued. We have the following result.\\

{\bf Theorem 5.5.} {\em Let $t$ be a positive integer satisfying $t \geq \frac{n}{2}$. We have an explicit  length $n$ subspace-metric code over ${\bf F}_{q^{2n}}$ with the subspace distance $2(n-t)$ and the cardinality $q^{\frac{3n}{2}(t+1)-\frac{n^2}{4}}\cdot \frac{4(q^{\frac{n}{2}}-1)}{n^2}$.}\\

Notice that the Singleton bound for a  length $n$ subspace-metric code with the insdel distance $2(n-t)$ is $q^{2n(t+1)}$.\\

 We refer the Levenshtein bound to \cite{L66,KLO}. It asserts  that $N(n, q, d)=\max\{|C| \subset {\bf F}_q^n: d_{insdel} >2 d\}$ satisfies $$N(n, q, 1) \leq \lfloor\frac{q^{n-1}+(n-2)q^{n-2}+q}{n}\rfloor.$$ This upper bound was improved to the upper bound $$N(4,q,1) \leq \frac{q^2(q+1)}{4}$$ in the case $n=4$ and $q$ even in \cite{KLO}. Some explicit length $4$ insdel codes over general alphabets attaining this bound was constructed in \cite{KLO}. An improvement on the Levenshtein upper bound was given in a recent paper \cite{Y21}. From the result in the previous section and the lower bound on the subspace codes in \cite{EtzionVardy,KSK} there exists a subspace-metric code over ${\bf F}_{q^n}$ with the length $u<<n$, the subspace distance $4$ and the size at least $q^{nu-u^2-3n+3u}$. Then there exists an $1$-deletion correcting code over ${\bf F}_{q^n}^u$ with the insdel distance $4$ and the size at least $q^{nu-u^2-3n+3u}$. The Levenshtein upper bound claims that the maximal possible size of $1$-deletion correcting code is $\lfloor\frac{q^{nu-n}+(u-2)q^{nu-2n}+q}{u}\rfloor$.  It is clear $1$-deletion correcting code from subspace coding has the size close to the Levenshtein upper bound.\\

\section{Explicit subspace-metric codes from orbit cyclic subspace codes}

A linear subspace $V$ over ${\bf F}_q$ in ${\bf F}_q^n={\bf F}_{q^n}$ is a Sidon space if for $a,b,c,d \in V$ satisfying $ab=cd$, then $\{a{\bf F}_q, b{\bf F}_q\}=\{c{\bf F}_q, d{\bf F}_q\}$. For each positive integer $k$ satisfying $k<\frac{n}{2}$, explicit $k$ dimension Sidon spaces were given in \cite{RRT}. These Sidon spaces were also used in \cite{CSI} for constructing explicit two dimensional Reed-Solomon codes attaining the half-Singleton bound.\\

A cyclic subspace code ${\bf C}$ is a subspace codes consisting of subspaces in ${\bf F}_{q^n}={\bf F}_q^n$, which is closed under the multiplication of non-zero elements of ${\bf F}_{q^n}$ on subspaces. That is if ${\bf C}$ is an $(n,M,d,k)_q$ constant dimension subspace code satisfying that for each $k$ dimension subspace $L \in {\bf  C}$, $xL$ as a $k$ dimension subspace in ${\bf F}_{q^n}$, for a non-zero element $x \in {\bf F}_{q^n}$, is also a codeword in ${\bf C}$. Explicit orbit cyclic constant dimension subspace codes were given in \cite{BEGR, Gluesing,RRT,CL17} from subspace polynomials and Sidon spaces. More explicitly, it is cyclic constant subspace code $${\bf C}=orbit(L)=\{xL:x \in {\bf F}_{q^n}, x\neq 0\}$$ where $L$ is a fixed $k$ dimension Sidon subspace in ${\bf F}_{q^n}$.  These orbit cyclic constant dimension subspace codes have the cardinality $|{\bf C}|=\frac{q^n-1}{q-1}$ and the maximal possible subspace distances $d_S({\bf C})=2k-2$.\\

{\bf Theorem 6.1} {\em Let $k$ and $n$ be two positive integers satisfying $k <\frac{n}{2}$. Then an explicit one dimension linear subspace-metric code $Span({\bf C})$ over ${\bf F}_{q^n}$ with the length $k$ and the subspace distance $d_S(Span({\bf C}))=2k-2$ can be given directly from the above orbit cyclic constant dimension subspace code ${\bf C}$.}\\

{\bf Proof.} From the construction in Theorem 6.1 the $Span({\bf C})$ has codeword of the form $x{\bf c}(L)$ for the Sidon space $L$. Then by adding one zero codewords we get the one dimension linear subspace-metric code.\\

The insdel distance of a one-dimension linear insertion-deletion code is close to the half-Singleton bound $d_{insdel}=2k-2 \leq 2(k-2+2)=2k$. \\

{\bf Corollary 6.1} {\em Let $k$ and $n$ be two positive integers satisfying $k <\frac{n}{2}$, we have an explicit $n$ dimension linear insertion-deletion code over ${\bf F}_q$ with the length $kn$ and insdel distance $2k-2$.}\\

{\bf Proof} The proof is similar to the proof of Theorem 5.3.\\

\section{$k$-deletion correcting codes with rate approaching $1$}

For a fixed positive integer  $k$, let $L_k(n)$ be the largest size of length $n$ binary codes correcting $k$-deletions, that is, $L_k(n)=\max \{|{\bf C}| \subset {\bf F}_2^n: d_{insdel}({\bf C}) >2k\}$, it was proved in \cite{L66} $$\frac{2^k(k!)^22^n}{n^{2k}} \leq L_k(n)\leq \frac{k!2^n}{n^k}.$$ The Varshamov-Tenengolts code can correct one deletion with the optimal redundancy $log(n+1)$. The problem to construct explicit $k$-deletion correcting codes of rate approaching $1$ as length grows had been unsettled for a long time. There have been continuous efforts in \cite{HF,PAGFC,BGZ,GS,SRB,SB,GH21} to construct length $n \longrightarrow \infty$ binary codes correcting $k$-deletions with optimal redundancies and the above long-standing problem was finally solved.  From Theorem 5.3 for given positive integer $k$ we can give a family of explicit $k$-deletion correcting codes with rate approaching $1$ immediately. This shows that our insdel-subspace-coding connection is a powerful method to give near-optimal insertion-deletion codes.\\

{\bf Theorem 7.1.} {\em Let $k$ be a fixed positive integer. $k$-deletion correcting code family with rate approaching to $1$ can be constructed from constant subspace codes directly.}\\

{\bf Proof.} We consider the case $k=2$ and $d_{insdel} \geq 6$. Then $(n, M, 6, u)_2$ constant dimension subspace codes can be used to construct a length $un$ binary code correcting $2$-deletions with the cardinality $M$ from Theorem 5.3. We will take $u<<n$, then from the lifted MRS code we have $M \geq 2^{(n-u)(u-6+1)}=2^{nu-u^2-5n+5n}$. Hence the rate $$R \geq \frac{nu-5n-u^2+5u}{nu}=1-\frac{5}{u}-\frac{u}{n}+\frac{5}{n}.$$ Then when $n$, $u$  and $\frac{n}{u}$ grow to the infinity, we have a family of binary $2$-deletion correcting codes with rate approaching  $1$. Similarly for fixed $k$ we have a family of length $un$ $k$-deletion correcting codes with the rate $$R\geq 1-\frac{u}{n}-\frac{2k+1}{u}+\frac{2k+1}{n}.$$  When $n$, $u$  and $\frac{n}{u}$ grow to the infinity, we have a family of binary $k$-deletion correcting codes with rate approaching  $1$. Since lifted MRD constant dimension subspace codes are explicitly given, our binary $k$-deletion codes are explicit codes. \\

\section{Subset distance of the folded codes from evaluation codes of linear mappings}

The folded code was introduced in \cite{GR08} for the purpose of achieving list-decoding capacity. The folded  codes are natural subset-metric codes and then insertion-deletion codes, as showed in the following example.\\

We consider the evaluation codes of all linear mappings $f: {\bf F}_q^n \longrightarrow {\bf F}_q$, on the set of some points ${\bf X}=(p_1,\ldots, p_n)$,  $p_1, p_2, \ldots, p_n \in {\bf F}_q^n$, may be repeated. This set ${\bf X}$ will be determined as follows.\\

Let ${\bf e}_1, \ldots, {\bf e}_n$ be linear independent elements in ${\bf F}_q^n={\bf F}_{q^n}$ over ${\bf F}_q$. For one such linear function, $(f({\bf e}_1),\ldots,f({\bf e}_n))$ is an element in ${\bf F}_{q^n}$. The element $(f({\bf x} {\bf e}_1), \ldots, f({\bf x} {\bf e}_n)) \in {\bf F}_{q^n}$ can be represented as the multiplication ${\bf x} (\Sigma_{i=1}^n f({\bf e}_i){\bf e}_i)$ of two elements ${\bf x}$ and $\Sigma_{i=1}^n f({\bf e}_i){\bf e}_i$. Actually if ${\bf x} \cdot {\bf e}_i=x_{i1}{\bf e}_1+\cdots+x_{in}{\bf e}_n \in {\bf F}_{q^n}={\bf F}_q^n$, $x_{ij} \in {\bf F}_q$, then $f({\bf x}\cdot  {\bf e}_i)=x_{i1}f({\bf e}_1)+\cdots+x_{in}f({\bf e}_n)$.\\

For another linear function $$g: {\bf F}_q^n \longrightarrow {\bf F}_q,$$ we can represent $g$ as $g({\bf z})=f({\bf y} \cdot {\bf x})$ for any ${\bf z} \in {\bf F}_{q^n}$, with a fixed non-zero element ${\bf y} \in {\bf F}_{q^n}$. For a subset ${\bf D} \subset {\bf F}_{2^n}^*$, the point set ${\bf X}$ is the set of all points in $({\bf x} \cdot {\bf e}_1, \ldots, {\bf x}\cdot {\bf e}_n)$ for all nonzero elements ${\bf x} \in {\bf D}$. This is the evaluation code ${\bf C}$ depending on the subset ${\bf D} \subset {\bf F}_{q^n}^*$ with the cardinality $|{\bf D}|=D$. We will analysis the subset distance of the folded code $Folded({\bf C})$ of this code ${\bf C}$ with the parameter $n$, that is, the codewords in $Folded({\bf C})$ is of the form $({\bf c}_1, \ldots, {\bf c}_D)$, where ${\bf c}_i=(f({\bf x}_i {\bf e}_1), \ldots, f({\bf x}_i {\bf e}_n))={\bf x}_i(\Sigma_{j=1}^n f({\bf e}_j){\bf e}_j)$, where $${\bf D}=\{{\bf x}_1, \ldots, {\bf x}_D\}.$$ For another linear function $g({\bf z})=f({\bf y} \cdot {\bf z})$, then the codewords is of the form $({\bf y} {\bf x}_1(\Sigma_{j=1}^n f({\bf e}_j){\bf e}_j), \ldots, {\bf y} {\bf x}_D(\Sigma_{j=1}^n f({\bf e}_j){\bf e}_j))$.\\

We define $$m({\bf D})=\max \{|{\bf y} \cdot {\bf D} \bigcap {\bf D}| : {\bf y } \in {\bf F}_{q^n}^*\}.$$ Then we have the following result.\\

{\bf Theorem 8.1.} {\em The subset distance of the folded code $Folded({\bf C})$ is $D-m({\bf D})$.}\\

{\bf Proof.} The conclusion follows from the representation of codewords in the folded code $Folded({\bf C})$.\\

A subset ${\bf D} \subset {\bf F}_{2^n}^*$ of the cardinality $k$ is called an $(2^n-1, k, \lambda)$-difference set if for every non-identity element ${\bf y} \in {\bf F}_{2^n}^*$, we have $$|{\bf y} \cdot {\bf D} \bigcap {\bf D}|=\lambda,$$ we refer to Chapter 4 of \cite{Ding}. From Theorem 8.1 for such a difference set we have an folded code with length $k$ and the subset distance $k-\lambda$. It is well-known there is the Singer difference sets with the parameter $(2^n-1,2^{n-1}-1, 2^{n-2}-1)$ for $n \geq 3$, we refer to page 97 in \cite{Ding}. Then we have a folded code of length $2^{n-1}-1$ and the subset distance $2^{n-2}$. The size of this folded code is $2^{n-2}$. Hence we have an insertion-deletion code over ${\bf F}_{2^n}$ of length $2^{n-1}-1$, the cardinality $2^{n-2}$ and the insdel distance at least $2^{n-2}$.\\

{\bf Corollary 8.1.} {\em Let $n$ be a positive integer satisfying $n \geq 3$. Based on the Singer classical difference set in the multiplicative Abelian group ${\bf F}_{2^n}^{*}$, we have an explicit folded code over ${\bf F}_{2^n}$ with the length $2^{n-1}-1$, the cardinality $2^{n-2}$ and the subset distance $2^{n-2}$ from the evaluation codes of linear mappings.}\\

\section{Encoding and decoding}

In our  above insertion-deletion codes constructed from subspace codes,  only lifted MRD codes and some well-constructed constant subspace codes are used. Hence the encoding and decoding are direct from the encoding and decoding of the corresponding Gabidulin codes and are both highly efficient. On the other hand most explicit constant dimension subspace codes given in \cite{table} are explicitly constructed from several block combining of lifted MRD codes. Hence most subspace-metric codes or insertion-deletion codes constructed in  Theorem 5.1, Theorem 5.2, Theorem 5.3, Theorem 6.1 and Corollary 6.1 have highly efficient encoding and decoding.

\section{Conclusion}

In this paper we introduce the subspace-metric codes and subset-metric codes. We show that subspace distance and subset distance are natural lower bounds for the insdel distance and prove the half-Singleton bounds for the subspace distances and subset distances of linear codes. From subspace-metric and subset-metric codes well-structured explicit good insertion-deletion codes correcting synchronization errors can be constructed directly. Explicit near-Singleton bound subspace-metric codes over large fields are constructed. Insertion-deletion codes from subspace codes are close to the Levenshtein upper bound for $1$-deletion correcting codes over general finite field.   $k$-deletion correcting codes with rate approaching $1$ can be constructed from subspace codes. The one-dimension linear subspace-metric codes are given explicitly from the orbit cyclic subspace code. This lead to a high dimension linear insertion-deletion code with a natural lower bound on its insdel distance. The highly nontrivial lower bound for the subset distances of some folded codes is proved. Further construction of explicit synchronization strings from subspace coding will be presented in \cite{Chen212}.\\

\end{document}